\documentclass{IEEEtran}
\usepackage{graphicx}
\usepackage{amsmath}
\usepackage{multicol}
\usepackage{multirow}
\usepackage{stfloats}
\usepackage{booktabs}
\usepackage{mathrsfs}
\usepackage{enumerate}
\usepackage{amssymb}
\usepackage{algorithm}
\usepackage{array}
\usepackage{underscore}
\usepackage{url}

\usepackage{hyperref}
\hypersetup{hidelinks}
\usepackage{flushend}

\usepackage{algorithmic}
\usepackage[square, comma, sort&compress, numbers]{natbib}
\usepackage{graphicx}
\usepackage{epstopdf}
\usepackage{caption}
\usepackage{diagbox}

\usepackage{caption}

\usepackage{subfigure}
\usepackage{color}

\usepackage[outerbars,color]{changebar}
\ifx\pdfoutput\undefined
\else\ifnum\pdfoutput>0
  \usepackage{pdfcolmk}
\fi\fi
\cbcolor{red}
\usepackage{fancyhdr}

\makeatletter
\newcommand{\Rmnum}[1]{\expandafter\@slowromancap\romannumeral #1@}
\makeatother

\usepackage{array}  \newcommand{\PreserveBackslash}[1]{\let\temp=\\#1\let\\=\temp}  \newcolumntype{C}[1]{>{\PreserveBackslash\centering}p{#1}}  \newcolumntype{R}[1]{>{\PreserveBackslash\raggedleft}p{#1}}  \newcolumntype{L}[1]{>{\PreserveBackslash\raggedright}p{#1}}

\ifCLASSINFOpdf
\else
\fi
\begin{document}
\title{\begin{huge} Wireless Energy Transmission Channel Modeling in Resonant Beam Charging for IoT Devices
\end{huge}}

\author{Wei~Wang\IEEEauthorrefmark{2}, 
Qingqing~Zhang\IEEEauthorrefmark{2},
Hua~Lin,
Mingqing~Liu,
Xiaoyan~Liang\IEEEauthorrefmark{1},
and Qingwen~Liu\IEEEauthorrefmark{1}


    \thanks{W.~Wang, H.~Lin and X. Liang, are with the State Key Laboratory of High Field Laser Physics, Shanghai Institute of Optics and Fine Mechanics, Chinese Academy of Sciences, Shanghai, China. W.~Wang is also with the University of Chinese Academy of Sciences, Beijing 100049, China, (email: wangwei2016@siom.ac.cn, hual@siom.ac.cn, liangxy@siom.ac.cn). }%
    \thanks{Q.~Zhang, M. Liu, and Q. Liu, are with the College of Electronic and Information Engineering, Tongji University, Shanghai, China, (email: anne@tongji.edu.cn, 18392105294@163.com, qliu@tongji.edu.cn). }%
    \thanks{$\dagger$ Both authors contributed equally to this work.}
	\thanks{* Corresponding author.}%
}
\maketitle

\begin{abstract}
Power supply for Internet of Things (IoT) devices is one of the bottlenecks in IoT development. To provide perpetual power supply to IoT devices, resonant beam charging (RBC) is a promising safe, long-range and high-power wireless power transfer solution. How long distance can RBC reach and how much power can RBC transfer? In this paper, we analyze the RBC's consistent and steady operational conditions, which determine the maximum power transmission distance. Moreover, we study the power transmission efficiency within the operational distance, which determines the deliverable power through the RBC energy transmission channel. Based on this energy transmission channel modeling, we numerically evaluate its impacts on the RBC system performance in terms of the transmission distance, the transmission efficiency, and the output electrical power. The analysis leads to the guidelines for the RBC system design and implementation, which can deliver multi-Watt power over multi-meter distance wirelessly for IoT devices.

\end{abstract}

\IEEEpeerreviewmaketitle

\section{Introduction}
With the development of the Internet of Things (IoT) \cite{huang2013iot,ling2017iot,ding2014iot} and the Big Data \cite{yu2017bigdata}, the power supply for IoT devices has become one of the bottlenecks of IoT development. However, carrying power cord and looking for power supply cause inconvenience for people. The contradiction between battery endurance of IoT devices and the power supply is increasingly prominent \cite{Carroll2010,Georgiou2017,Yu2016A,zhang2017networks}. Therefore, wireless power transfer (WPT) draws much attention to provide mobile power supply anywhere and anytime for IoT devices.

Several kinds of WPT technologies have been well investigated in research \cite{wirelesstechniques,electromagnetic}. Inductive coupling \cite{inductive} is safe and simple but limited by a short charging distance from a few millimeters to centimeters. Magnetic resonance coupling \cite{magnetic,Cannon2009} has high charging efficiency, however it is restricted by a short charging distance and a large coil size. Radio frequency (RF) \cite{RF} has a long effective charging distance. However, it suffers low efficiency and difficulty to balance safety and high power. Laser charging \cite{Lasercharging,Sahai2011} needs directional pointing and can transmit high power to a long distance. But it also faces the radiation safety challenge. In summary, these technologies face with technical challenges to satisfy safety, long-range, high-power at the same time.

Resonant beam charging (RBC), as known as distributed laser charging (DLC), is presented in \cite{liu2016dlc} which can safety provide multi-Watt wireless power supply over multi-meter distance for IoT devices. In the RBC system, any object blocking the line of sight (LOS) between the transmitter and the receiver can break resonation immediately, which leads to the inherent safety. Meanwhile when the RBC system operates as a stable resonant cavity, even slight blocking of the beam will instantly cause a significant alteration in the beam's power, which makes it simple to detect such a blocking. Furthermore, any bystander outside of the resonant cavity limits will not be subjected to any of the beam's radiation. The adaptive resonant beam charging (ARBC) system proposed in \cite{Qing2017} improved the WPT efficiency based on feedback control. Moreover, the RBC not only can achieve self-aligning, but also can charge multi-devices simultaneously.

\begin{figure}
	\centering
	\includegraphics[width=3.0in]{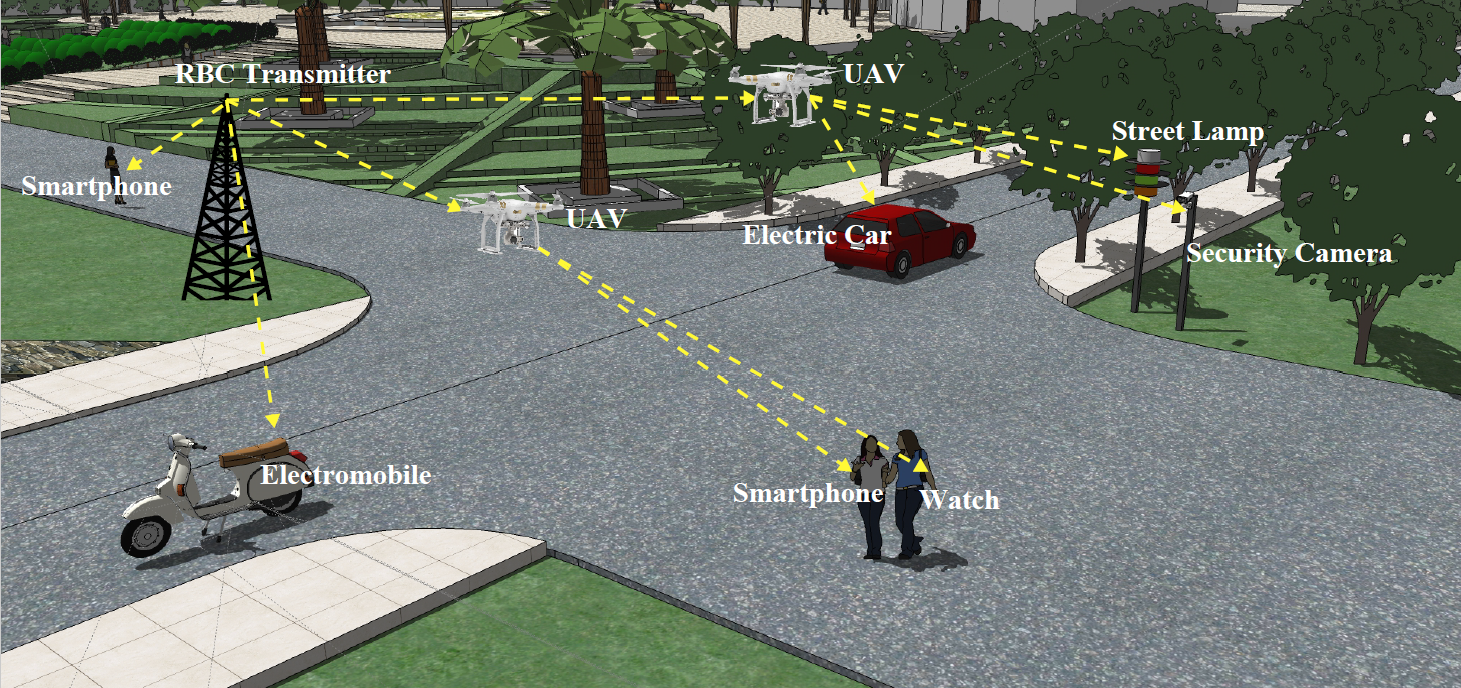}
	\caption{Resonant Beam Charging Application}
	\label{application}
\end{figure}

Fig.~\ref{application} illustrates the RBC potential application. As shown in Fig.~\ref{application}, the RBC transmitter can be mounted on a communication base station, which can provide wireless power to electromobile and mobile devices within its coverage. In addition, the unmanned aerial vehicle (UAV) equipped with both the RBC transmitter and receiver can play the role of a relay to receive power from the RBC transmitter and transmit power to IoT devices within its coverage.
The RBC is a promising solution to break the power supply bottleneck in development of IoT. However, how long distance RBC can reach and how much power RBC can transfer are essential issues to investigate. The contributions of this paper include: 1) We present the analytical model of the energy transmission channel for the RBC system and study the RBC's consistent and steady operational conditions and the power transmission efficiency within the operational distance;
2) We numerically evaluate its impacts on the RBC system performance in terms of the transmission distance, the transmission efficiency, and the output electrical power;
3) We derive the closed-formula for the relationship among the end-to-end power, efficiency and transmission distance, which provide guidelines for the RBC system design and implementation.

\begin{figure}
	\centering
	\includegraphics[width=3.0in]{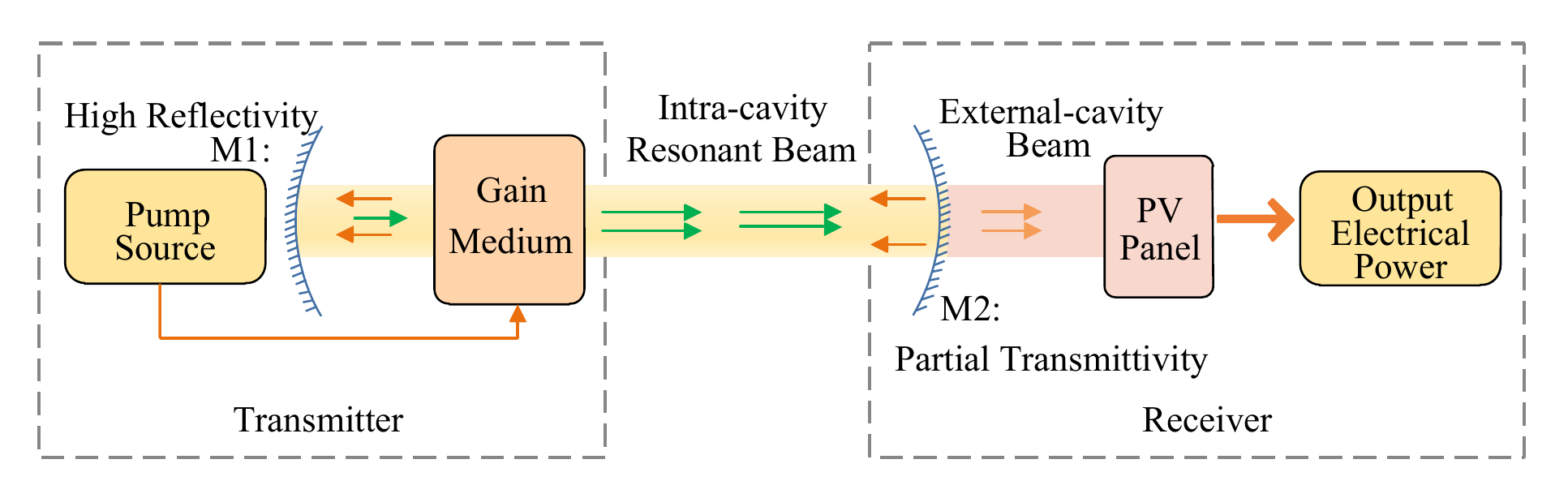}
	\caption{Resonant Beam Charging System}
	\label{fig:system-modle}
\end{figure}	

In the rest of this paper, we will briefly introduce the RBC system at first. Then, we will give the modular model and working mechanism of the RBC system. Based on the analytical analysis, we will depict the system performance in the next section. Finally, we will make a conclusion and discuss open issues for future research.


\section{Resonant Beam Charging}

The resonant beam charging (RBC) is a WPT technology which transfers power through intra-cavity resonant beam. In the RBC system, the optical components are divided into two separate parts: the transmitter and the receiver. The architecture of the RBC system is shown in Fig.~\ref{fig:system-modle}. In the RBC transmitter, there is a high reflectivity curved mirror M1 and a gain medium pumped by pump source. The pump source excites gain medium to realize population inversion, which leads to energy storage in the gain medium. The energy stored in the gain medium is determined by the input electrical power, the performance of pump source, the pump chamber structure, and the property of gain medium \cite{SolidLaser}.

While in the RBC receiver, a partial transmittivity curved mirror M2 is contained. M1 and M2 constitute the optical resonant cavity, which performs the function of a highly selective feedback element by coupling a portion of the signal generated from the gain medium in phase. The resonant cavity should satisfy steady condition for providing sufficient large feedback for the RBC system. At the same time, satisfy steady condition can ensure high safety and efficiency for the resonant cavity. If the feedback is large enough to compensate the internal losses of the system, the system starts to oscillate. When the system is triggered by the spontaneous radiation emitted along the axis of the resonant cavity. Thus, the gain medium stored energy is released to the intra-cavity resonant beam by stimulated emission \cite{SolidLaser}. The transmission efficiency relies on the energy transmission distance of the system and the loss related to the transverse mode.

The beam emitted by most optical resonant cavity contains several discrete optical frequencies. The beams are separated from each other due to frequency differences which can lead to different modes of the optical resonant cavity. Each mode is defined by the variation of the electromagnetic field perpendicular and along the axis of the resonant cavity. It is common to distinguish two types of resonant cavity modes: longitudinal modes which differ from each other only in their oscillation frequency; transverse modes which differ from each other not only in their oscillation frequency, but also in their field distribution in a plane perpendicular to the direction of propagation \cite{SolidLaser}.

The loss affects the gain medium stored energy to intra-cavity beam power transmission efficiency. There are two kinds of loss with different properties in the RBC system \cite{diffraction}. The first kind is the loss independent of the resonant beam transverse mode, such as the internal loss of the gain medium, the transmission loss of the mirror, the absorption, and scattering loss of the element in the system. The other is the diffraction loss that closely related to the transverse mode and the transmission distance.

At the same time in the RBC receiver, a photovoltaic panel (PV-panel) is installed behind the mirror M2. The power of intra-cavity resonant beam is extracted from the partial transmittivity mirror M2. The external-cavity beam power can be converted to electrical power by the PV-panel. The beam power-to-electrical power conversion is based on the PV engineering \cite{secondauthorPV,green2015solar,Summerer2008Concepts}.

\section{Analytical Modeling for Energy Transmission}
\begin{figure}
	\centering
	\includegraphics[width=3.0in]{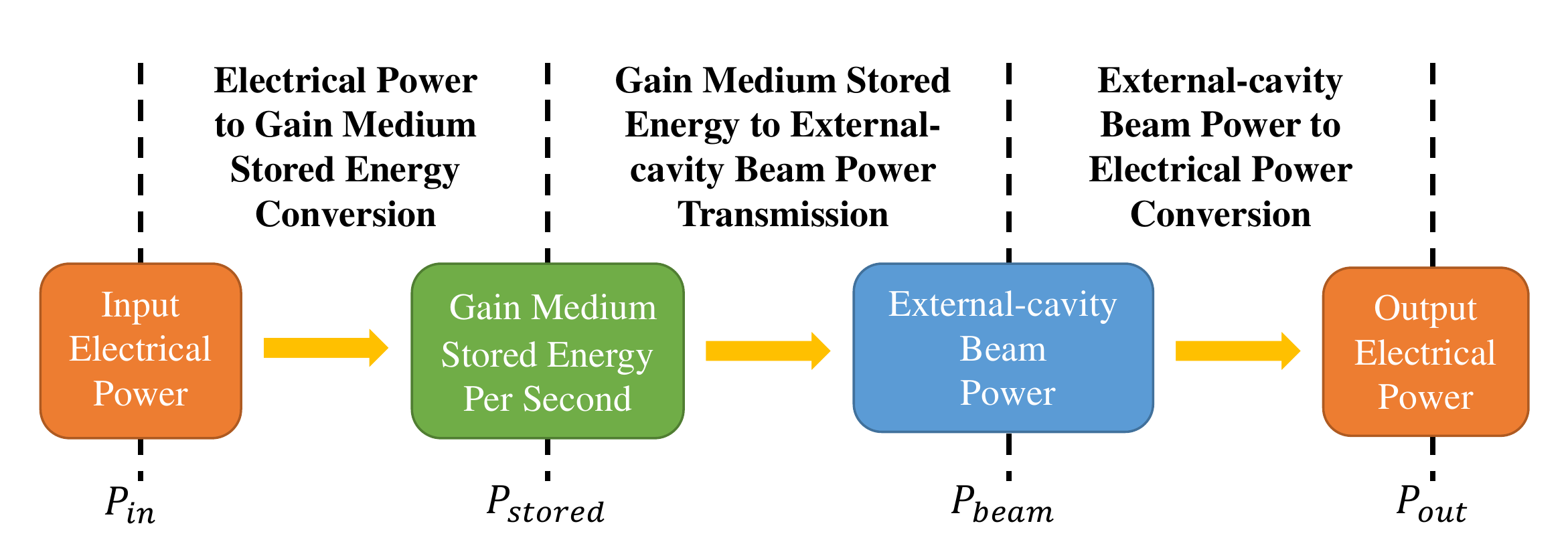}
	\caption{Modular Model of the RBC System}
	\label{fig:electricity-electricity}
    \end{figure}
In the RBC system, the input electrical power will be converted to the gain medium stored energy. Then, the gain medium stored energy will be converted to the intra-cavity resonant beam power, which will be transmitted from the transmitter to the receiver over a certain distance. Then, the intra-cavity resonant beam power will be partially converted to the external-cavity beam power at the receiver. Finally, the external-cavity beam power will be converted to the output electrical power, which can be used to charge devices accessed to the RBC system.

In this section, we will illustrate the analytical modeling of the RBC energy transmission channel by dividing the procedure into three stages: the input electrical power to the gain medium stored energy conversion, the gain medium stored energy to the external-cavity beam power transmission, and the external-cavity beam power to the output electrical power conversion. Fig.~\ref{fig:electricity-electricity} gives the modular model and the three energy transmission stages of the RBC system.

\subsection{Electrical Power-to-Stored Energy Conversion}

The input electrical power ${P_{in}}$, which is provided by the driving source, acts on the pump source. The pump source transforms the electrical power into the pump power which can be absorbed by the gain medium. Then, the pump power is transferred to the gain medium through a completely enclosed reflective chamber i.e. pump chamber. The absorbed pump power excites the gain medium to realize the population inversion, and then the energy is stored in the gain medium.

In the case of continuous input of electrical power, the relationship between the gain medium stored energy per second ${P_{stored}}$ and the input electrical power ${P_{in}}$ can be depicted as \cite{SolidLaser}:
     \begin{equation}\label{gs-pstorepin}
    {P_{stored}} = {\eta _{stored}}{P_{in}},
    \end{equation}
where $\eta _{stored}$ is the electrical power-to-stored energy conversion efficiency. It is affected by the performance of the pump source, the pump chamber structure, and the material and size of gain medium \cite{SolidLaser}. ${\eta _{stored}}$ can be computed as:
    \begin{equation}\label{gs-erastim}
    {\eta _{{\rm{stored}}}} = \frac{{{P_{stored}}}}{{{P_{in}}}}.
    \end{equation}
Most of the input electrical power which can not be converted to the stored energy is dissipated as heat.

\subsection{Stored Energy-to-Beam Power Transmission}
The stored energy at the transmitter can be triggered by the spontaneous radiation emitted along the axis of the resonant cavity, and be converted to the intra-cavity resonant beam power. After part of the intra-cavity resonant beam passing through the partial transmittivity mirror M2, the external-cavity beam is formed. Therefore, with the transmission attenuation in the free space, the stored energy will be converted into the external-cavity beam power. Since both the transmission distance and the transmission efficiency have effects on the stored energy to beam power transmission efficiency, we will give the transmission distance model at first. Then, we will analyze the factors which influence the transmission efficiency.

\subsubsection{Transmission Distance Modeling}
To support the energy transmission, the resonant cavity should keep stable at first. Factors that influence the stable condition include the material and construction of the gain medium, the curvature radius of the mirror at the transmitter M1 and that of the mirror at the receiver M2, the distance between the gain medium and M1 $l$, and the distance between the gain medium and M2 $d$.

\begin{figure}
	\centering
	\includegraphics[width=3.0in]{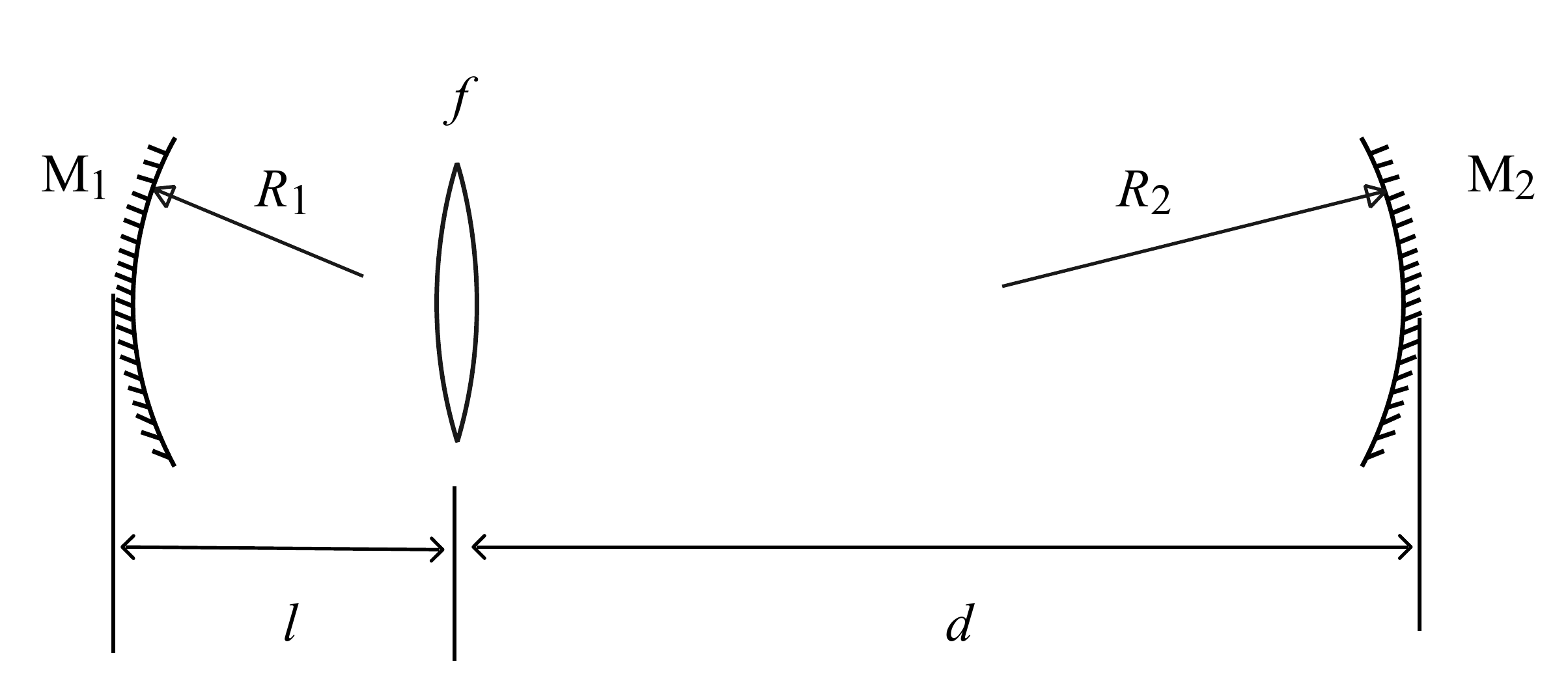}
	\caption{Schematic Diagram of Resonant Cavity with Thermal Lens}
	\label{fig:cavity}
\end{figure}

A resonant cavity containing a rod shaped gain medium is taken as an example. In the rod, the heat generation is uniform, which acts as a lens-like medium since, due to thermal and stress-induced effects. From \cite{LENS}, the resonant cavity can be represented by a model containing a thin lens, as shown in Fig.~\ref{fig:cavity}. Where the gain medium is replaced by a thin lens with a focal length of $f$, $R1$ is the curvature radius of M1 mirror, $R2$ is the curvature radius of M2 mirror, $l$ represents the space size of the transmitter, $d$ represents the transmission distance. We define the effective resonant cavity length $L$ as:
    \begin{equation}
    L = {l} + {d} - \frac{{{l}{d}}}{f}.
    \label{equ:L}
    \end{equation}

To express the stability condition of the resonant cavity more clearly, we introducing $g_1$, $g_2$ as:
    \begin{equation}
    {g_1} = 1 - \frac{{{d}}}{f} - \frac{{L}}{{{R_1}}},
    \label{equ:g1}
    \end{equation}
    \begin{equation}
    {g_2} = 1 - \frac{{{l}}}{f} - \frac{{L}}{{{R_2}}}.
    \label{equ:g2}
    \end{equation}
As can be seen, $g_1$ and $g_2$ are both related to the transmission distance $d$. According to \cite{stable}, to keep the RBC system working at the stability condition, the following condition should be satisfied:
    \begin{equation}
    0 < {g_1}{g_2} < 1.
    \label{equ:g1g2}
    \end{equation}

\begin{figure}
	\centering
	\includegraphics[scale=0.27]{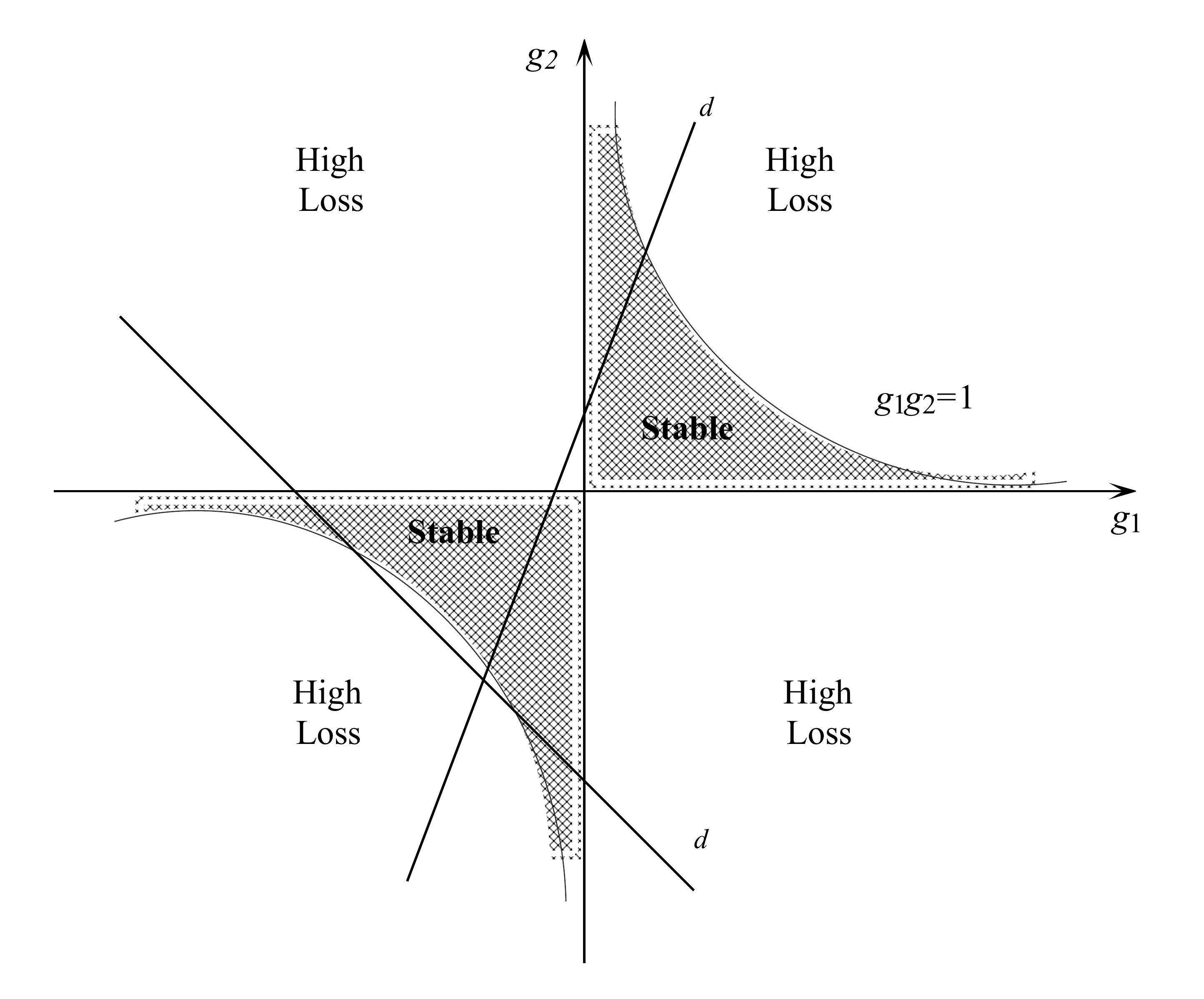}
	\caption{Stability Diagrams}
	\label{stabilitytheory}
\end{figure}

To show graphically which type of resonant cavity is stable and which is unstable, it is useful
to plot a stability diagram Fig.~\ref{stabilitytheory}. In Fig.~\ref{stabilitytheory}, each particular resonant cavity geometry is represented by a point to show the stable or unstable resonant cavity type. All cavity configurations are unstable unless they correspond to points located in the area enclosed by a branch of the hyperbola $g_1g_2=1$ and the coordinate axes, as shown in the shadow part of Fig.~\ref{stabilitytheory}. As the distance $d$ varies, the straight line in Fig.~\ref{stabilitytheory} is derived by eliminating $d$ in \eqref{equ:g1} and \eqref{equ:g2}:
    \begin{equation}
    {g_2} = \frac{{({l} - f){R_1}}}{{({l} - {R_1} - f){R_2}}}{g_1} + \frac{{({l} - f)({l} - {R_1})}}{{({l} - {R_1} - f){R_2}}} + 1 - \frac{{{l}}}{f} - \frac{{{l}}}{{{R_2}}}.
    \label{equ:g1g2}
    \end{equation}
The two straight lines in Fig.~\ref{stabilitytheory} correspond to two general situations. The intersections of the straight line with the axes and with the hyperbola $g_1g_2=1$ gives the critical values of $d$ corresponding to the edges of the stability regions \cite{stable0}. In our theoretical model, the range of $d$ satisfying the stability condition represents the distance range that can realize energy transmission. The critical value of $d$ determines the maximum transmission distance of the RBC system.

Though the intensity distribution on each section of the resonant beam is consistent, the beam radius (defined as the radius at which the electric field amplitude is down by $1/e$ from the maximum) of the intensity distribution varies along the optical axis \cite{SolidLaser}. For analyzing the relationship between the beam radius of each location and the transmission distance more concisely, we introduce three new variables $u_1$, $u_2$, and $x$ as:
\begin{equation}
{u_1} = {l}\left( {1 - \frac{{{l}}}{{{R_1}}}} \right),
\end{equation}
\begin{equation}
{u_2} = {d}\left( {1 - \frac{{{d}}}{{{R_2}}}} \right),
\end{equation}
\begin{equation}
x = \frac{1}{f} - \frac{1}{{{l}}} - \frac{1}{{{d}}}.
\end{equation}
The beam radius on gain medium $\omega _1$, on M1 $\omega _2$ and on M2 $\omega _3$ can be expressed as \cite{stable0}:
\begin{equation}
{\omega _{\rm{1}}}{\rm{ = }}\sqrt {\frac{\lambda }{\pi }\frac{{\left| {2x{u_1}{u_2} + {u_1} + {u_2}} \right|}}{{{{\left[ {\left( {1 - {g_1}{g_2}} \right){g_1}{g_2}} \right]}^{{1 \mathord{\left/
 {\vphantom {1 2}} \right.
 \kern-\nulldelimiterspace} 2}}}}}},
 \label{omega12}
 \end{equation}
  \begin{equation}
{\omega _2}{\rm{ = }}\sqrt {\frac{{\lambda \left| L \right|}}{\pi }{{\left[ {\frac{{{g_2}}}{{{g_1}\left( {1 - {g_1}{g_2}} \right)}}} \right]}^{{1 \mathord{\left/
 {\vphantom {1 2}} \right.
 \kern-\nulldelimiterspace} 2}}}},
 \label{omega22}
 \end{equation}
 \begin{equation}
{\omega _3}{\rm{ = }}\sqrt {\frac{{\lambda \left| L \right|}}{\pi }{{\left[ {\frac{{{g_1}}}{{{g_2}\left( {1 - {g_1}{g_2}} \right)}}} \right]}^{{1 \mathord{\left/
 {\vphantom {1 2}} \right.
 \kern-\nulldelimiterspace} 2}}}}.
 \label{omega32}
 \end{equation}

However, due to losses of the RBC system, the maximum transmission distance can not be actually obtained. There are two kinds of loss influencing the transmission: the loss independent of the beam transverse mode, and the diffraction loss closely related to the transverse mode \cite{diffraction}. The loss independent of transmission mode is determined by the intrinsic nature of the components used by the system. Therefore, we only take the transverse mode related diffraction loss into consideration.

\subsubsection{Transmission Efficiency Modeling}
\begin{figure}
	\centering
	\includegraphics[width=3.0in]{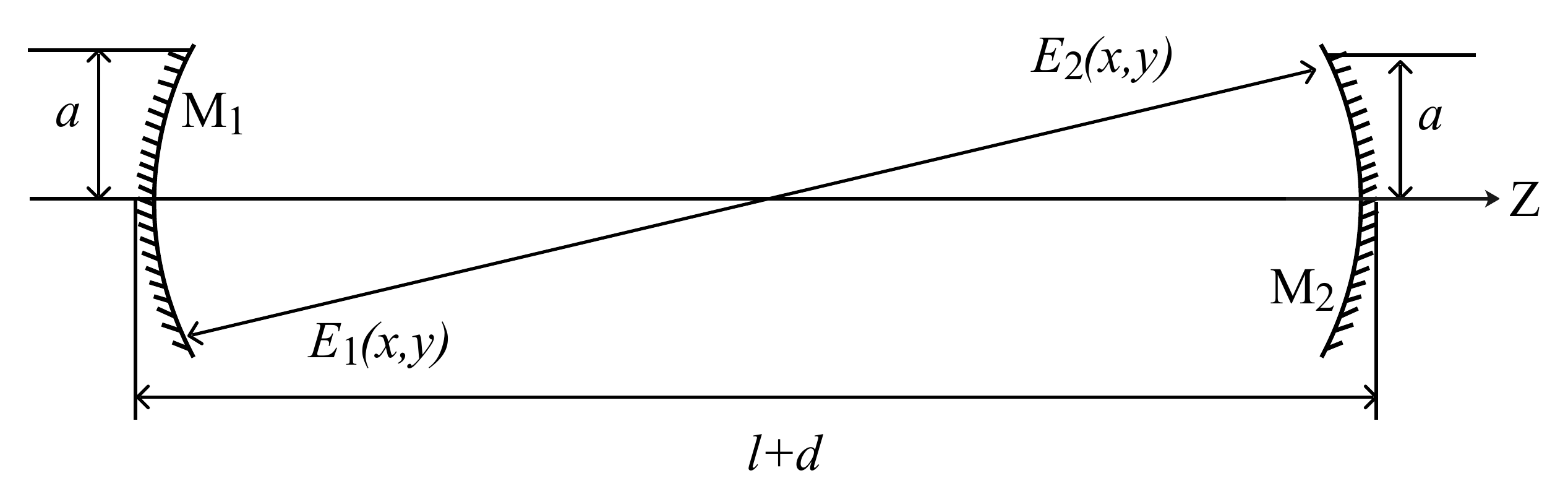}
	\caption{Schematic Diagram of Stable Resonant Cavity}
	\label{laser-resonantor0}
\end{figure}

The diffraction loss is mainly caused by the diffraction effect. According to \cite{Laser}, resonant beam is a beam oscillation with the feedback of the resonant cavity and the gain medium. When the resonant beam travels back and forth between the M1 and M2, the edge of the mirror M1 and M2 will cause loss due to the diffraction effect. Because the mirror geometry of M1 and M2 is limited, as shown in Fig.~\ref{laser-resonantor0}, where $a$ presents the radius of the end mirror M1 and M2 of the optical resonant cavity.
As shown in the Fig.~\ref{laser-resonantor0}, after a number of round-trip propagation, the electric field distributions of the cross section of a resonant beam on the mirror M1 and M2 are $E_1(x,y)$, and $E_2(x,y)$. Then, there should be the following self consistent relationship \cite{Laser}:
    \begin{equation}
    {E_2}(x,y) = \gamma {E_1}(x,y),
    \end{equation}
    where $\gamma$ is a complex constant factor, which reflects the change of amplitude and phase of beam field after a one-way propagation. The total energy loss of single trip $\delta$ can be expressed as:
    \begin{equation}
    {\delta} = \frac{{{{\left| {{E_1}} \right|}^2} - {{\left| {{E_2}} \right|}^2}}}{{{{\left| {{E_1}} \right|}^2}}} = 1 - {\left| \gamma  \right|^2}.
    \end{equation}

The loss which independent of the beam transverse mode, such as transmission, absorption and output, is depending on the intrinsic properties of the components. Therefore this loss is not considered in our model. Thus $\gamma$ only contains the diffraction loss of the edge of the mirror M1 and M2. Meanwhile, the diffraction loss is caused not only by the optical resonant cavity end mirrors, but also by finite aperture in the resonant cavity, such as the aperture of gain medium. According to \cite{Loss}, the diffraction loss of the beam through a aperture with an diameter of $2a$ in the optical resonant cavity $\delta_{mn}$ is:
    \begin{equation}
    {\delta _{mn}} = 1 - \frac{{\int\limits_0^{2\pi } {\int\limits_0^a {{r^{2m}}{{\left[ {L_n^m(\frac{{2{r^2}}}{{{\omega ^2}}})} \right]}^2}{e^{ - 2\frac{{{r^2}}}{{{\omega ^2}}}}}rdrd\varphi } } }}{{\int\limits_0^{2\pi } {\int\limits_0^\infty  {{r^{2m}}{{\left[ {L_n^m(\frac{{2{r^2}}}{{{\omega ^2}}})} \right]}^2}{e^{ - 2\frac{{{r^2}}}{{{\omega ^2}}}}}rdrd\varphi } } }},
    \end{equation}
where $m$,$n$ are positive integers, $r$ is the radial radius, $\varphi$ is phase difference, and \(L_n^m(\xi )\) is $n$ order associative Laguerre polynomials.

For the general stable optical resonant cavity, the equivalent principle of the confocal resonator can be used to transform the diffraction loss \cite{Laser}. In the RBC system, to improve the uniformity of beam cross section energy distribution, the resonant beam is the superposition of $TEM_{00}$ mode and multiple higher order modes. The $TEM_{00}$ presents the fundamental mode in a optical resonant cavity, which is inevitable in resonant cavity. Furthermore, the high-order mode is with large number, and being difficult to be analyzed. Since the diffraction loss of the higher order mode is proportional to the $TEM_{00}$ mode, we only take the diffraction loss of $TEM_{00}$ mode into consideration in this paper. The diffraction loss $\delta _{00}(d)$ of $TEM_{00}$ mode, which is related with $d$, can be presented as:
\begin{equation}
\delta _{00}(d) = e^{ - 2\pi \frac{{{a^2}}}{{\lambda (l+d)}}}.
\label{delta00}
\end{equation}

To sum up, given the gain medium and certain structure of the resonant cavity, the external-cavity beam power $P_{beam}$ can be stimulated by the stored power $P_{stored}$. The relationship between $P_{beam}$ and $P_{stored}$ can be depicted as \cite{SolidLaser}:
\begin{equation}\label{gs-pbeampstore}
 P_{beam} = f(d)P_{stored} + C,
\end{equation}
where $f(d)$ is the equation related to the transmission distance $d$, and $C$ is a constant which depends on the internal parameters of the system.

With reference to \eqref{delta00} and \cite{SolidLaser}, $f(d)$, which is closely related with the diffraction loss ${\delta_{00}}$, can be computed as:
\begin{equation}\label{gs-fddelta}
\begin{aligned}
f(d)
&= \frac{{2\left( {1 - R} \right)m}}{{\left( {1 + R} \right){\delta _{00}}\left( d \right) - \left( {1 + R} \right)\ln R}}\\
&= \frac{{2\left( {1 - R} \right)m}}{{\left( {1 + R} \right){e^{ - 2\pi \frac{{{a^2}}}{{\lambda (l+d)}}}} - \left( {1 + R} \right)\ln R}},
\end{aligned}
\end{equation}
where $R$ is the reflectivity of the output mirror M2, and $m$ is the overlap efficiency. 

Therefore, we can obtain the stored energy to beam power transmission efficiency $\eta _{trans}$:
\begin{equation}\label{gs-etatrans}
\eta _{trans} = \frac{P_{beam}}{P_{stored}} = f(d)+\frac{C}{P_{stored}}.
\end{equation}

\subsection{Beam Power-to-Electrical Power Conversion}
At the RBC receiver, the external-cavity beam power can be converted to electrical power, which can be used to charge devices accessed to the RBC system. For certain input beam power, the output electrical power will be different if with different load in the circuit. To obtain the maximum beam power to electrical power conversion efficiency, the PV-panel should work at the maximum output power state with the assistance of maximum power point tracking (MPPT) technology. As stated in \cite{Qing2017}, the maximum PV-panel output power $P_{pv}$ takes a linear relationship with the external-cavity beam power $P_{beam}$, which is as:
\begin{equation}\label{gs-pmpppbeam}
P_{pv} = a_1 P_{beam} + b_1.
\end{equation}

Therefore, the beam power-to-electrical power conversion efficiency, i.e. the PV-panel conversion efficiency, $\eta_{pv}$ depends on $P_{pv}$ and $P_{beam}$, which can be depicted as:
\begin{equation}\label{gs-etapv}
\eta_{pv} = \frac{P_{pv}}{P_{beam}} = a_1 + \frac{b_1}{P_{beam}}.
\end{equation}

In summary, the PV-panel converts the received beam power $P_{beam}$ to the output electrical power $P_{pv}$ with the conversion efficiency $\eta_{pv}$.

\subsection{End-to-end Power Transmission}

Based on the above theoretical analysis for each procedure, the end-to-end power relationship, that is the relationship between the input electrical power and the output electrical power, can be depicted as:
\begin{equation}\label{pe2e}
\begin{aligned}
P_{out}
&= a_1 [f(d){P_{stored}} + C] + b_1\\
&= a_1 f(d)\eta _{stored}P_{in} + a_1C + b_1.
\end{aligned}
\end{equation}

As can be seen, the system output power $P_{out}$ is affected by the input power $P_{in}$ and the transmission distance $d$ at the same time. Based on \eqref{pe2e}, the relationship between the end-to-end transmission efficiency $\eta_{all}$ and the input electrical power can be obtained as:
\begin{equation}\label{etae2e}
\begin{aligned}
\eta_{all}
&= \eta _{stored}\eta _{trans}\eta_{pv}\\
&= a_1\eta _{stored}\eta _{trans} + \frac{b_1}{P_{in}}\\
&= a_1\eta _{stored}f(d) + \frac{a_1C+b_1}{P_{in}}.
\end{aligned}
\end{equation}
As can be seen, $\eta_{all}$ is affected by both $P_{in}$ and $d$.

The transmission or conversion efficiency of each module and the RBC power transmission efficiency are listed in Table~\ref{tbl:efficiency}.

The analytical modeling of the RBC energy transmission is illustrated, and the factors influencing the energy transmission are analyzed in this section. On this basis, the performance evaluation of the RBC system will be presented in the next section.
\begin{table}[b]
\centering
\caption{Transmission or Conversion Efficiency}
\begin{tabular}{C{1.5cm} C{6cm}}
\hline
 \textbf{Parameter} & \textbf{Description}  \\
\hline
\bfseries$\eta _{stored}$ & electrical power to gain medium stored energy conversion efficiency \\
\bfseries$\eta _{trans}$ & gain medium stored energy to external-cavity beam power transmission efficiency \\
\bfseries$\eta_{pv}$ & external-cavity beam power to electrical power conversion efficiency \\
\hline
\bfseries$\eta_{all}$ & end-to-end transmission efficiency \\
\hline
\label{tbl:efficiency}
\end{tabular}
\end{table}


\section{Performance Evaluation}
Based on the analytical modeling in the previous section, we can find that the RBC system efficiency varies with the input electrical power, the electrical power to stored energy conversion efficiency, the material and size of the gain medium, the transmission distance, the reflectivity of the mirrors, the overlap efficiency, the diffraction loss of the intra-cavity beam, and the beam power to the electrical power conversion efficiency. In this section, we will give the numerical evaluation of the RBC system in certain scenarios. All the numerical evaluations are implemented in MATLAB and Simulink.
\begin{figure}
	\centering
	\includegraphics[scale=0.6]{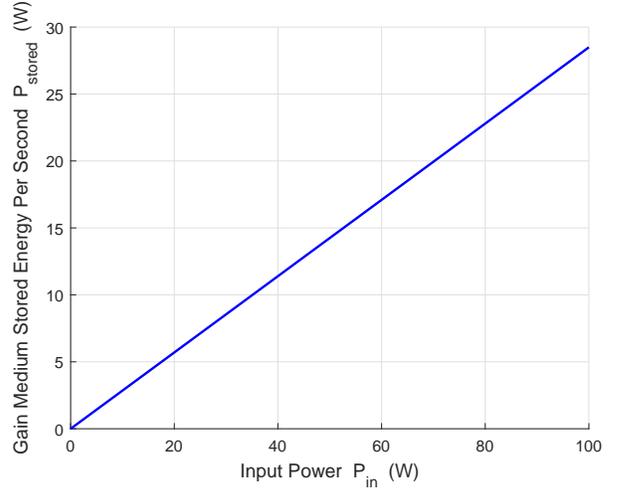}
	\caption{Gain Medium Stored Energy Per Second vs. Input Power}
	\label{pstorepin}
\end{figure}
\subsection{Electrical Power-to-Stored Energy Conversion}
At the RBC transmitter, the input electrical power $P_{in}$ is converted to the gain medium stored energy per second $P_{stored}$. The conversion efficiency is determined by the performance of pump source, the structure of the pump chamber and the material and size of the gain medium. Here we use the $808nm$ laser diode as the pump source, and the Nd:YAG rod, which is side pumped by the laser diode, as the gain medium. For the rod, the length is $65mm$, the diameter is $3mm$, and the stimulated beam is of $1064nm$. According to \cite{SolidLaser}, the electrical power to stored energy conversion efficiency $\eta_{stored}$ takes $28.49\%$.

Then, with reference to \eqref{gs-pstorepin}, the relationship between $P_{stored}$ and $P_{in}$ can be depicted out, as shown in Fig.~\ref{pstorepin}. As can be seen, $P_{stored}$ goes up linearly as $P_{in}$ increases, and the straight line between $P_{stored}$ and $P_{in}$ passes through the origin.

\subsection{Stored Energy-to-Beam Power Transmission}
In this part, we will illustrate how to design the cavity structure and select the transmission distance, and how the transmission distance influences the stored energy to beam power transmission efficiency.

\subsubsection{Transmission Distance Modeling}
According to \eqref{equ:L}, \eqref{equ:g1}, \eqref{equ:g2}, and \eqref{equ:g1g2}, given resonant cavity configuration, i.e. given $l$, $f$, $R_1$, and $R_2$, the position of a straight line in Fig.~\ref{stabilitytheory} is determined. Usually the focal length $f$ for Nd materials ranges from a few meters to a few tens of centimeters \cite{SolidLaser}. We set up $f=880mm$ according to the results of experimental measurements. According to the geometric parameters of the gain medium used in practice, the shortest $l$ is $60mm$. Usually we can obtain two distinct stability zones as shown in Fig.~\ref{stabilitytheory}. In an ideal wireless power transmission device, it is not desirable to have a range of distances that power can not be transferred within the maximum distance allowed. Thus we expect the transmission range to be continuous, that is, the two stability regions are connected.

\begin{figure}
	\centering
	\includegraphics[scale=0.6]{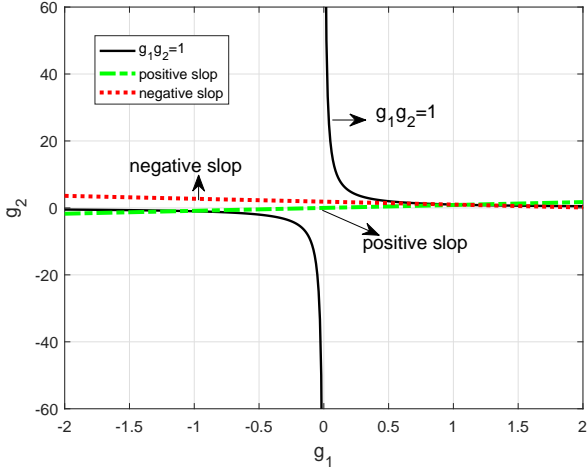}
	\caption{Connected Stability Diagrams}
	\label{stablezone}
\end{figure}
\begin{figure}
	\centering
	\includegraphics[scale=0.6]{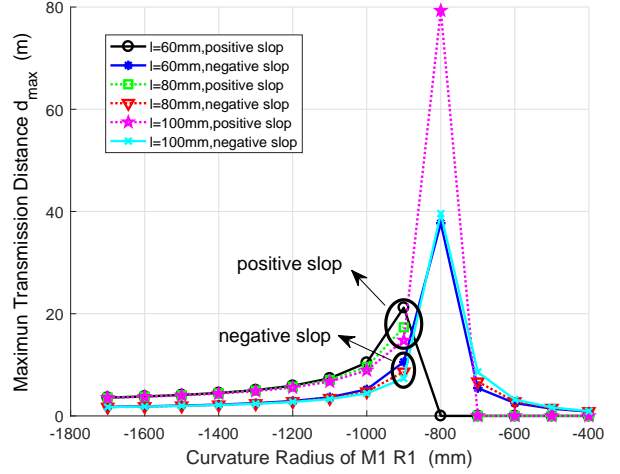}
	\caption{Maximum Transmission Distance vs. Curvature Radius of M1}
	\label{fig:transdistance}
\end{figure}
\begin{figure}
	\centering
	\vspace{-5mm} \includegraphics[scale=0.6]{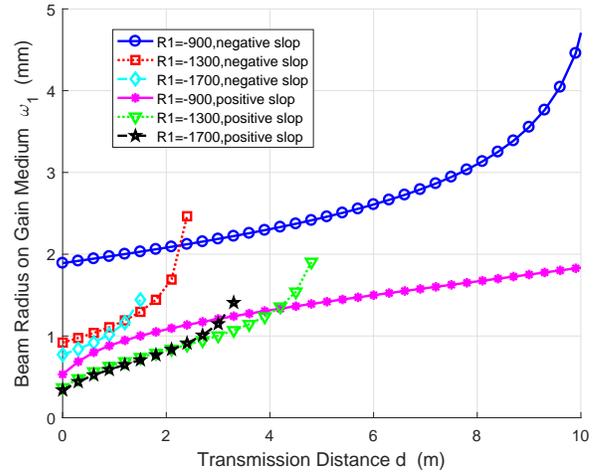}
	\caption{Beam Radius on Gain Medium vs. Transmission Distance}
	\label{fig:beammedium}
\end{figure}

To connect the two stability regions, the position of a straight line in Fig.~\ref{stabilitytheory} needs satisfy the intersection point of the straight line and the coordinate axis is coincident or the intersection point of the straight line and the $g_1g_2=1$ curve is coincident. There are two solutions: 1) the straight line slope is positive, and through the origin; 2) the straight line slope is negative and tangent to $g_1g_2=1$. Fig.~\ref{stablezone} shows the two situations that can connect two stability regions. The dashed presents the line of positive slope and the dotted line presents the line of negative slope. We can find that every set of values of $R1$ and $l$ has two $R2$ solutions, which is corresponding to two cases in Fig.~\ref{stablezone}.

No matter which cases the system works at, given $l$ and $R1$, the maximum intra-cavity resonant beam transmission distance $d_{max}$ can be different. The relationships among $l$, $R1$ and $d_{max}$ in the two cases are given in Fig.~\ref{fig:transdistance}. From Fig.~\ref{fig:transdistance}, when $R1=-800mm$, the solution of the negative slope about maximum transmission distance reaches the maximum as about $40m$. The maximum of positive slop is about $80m$. When $R1<-900mm$, the solution of positive slope is larger than that of the negative slope about maximum transmission distance. When $l$ takes $60mm$, $80mm$ and $100mm$, the maximum transmission distance is slightly different. In order to make the volume of the transmitter as small as possible, we set up $l=60mm$.
\begin{figure}
	\centering
	\includegraphics[scale=0.6]{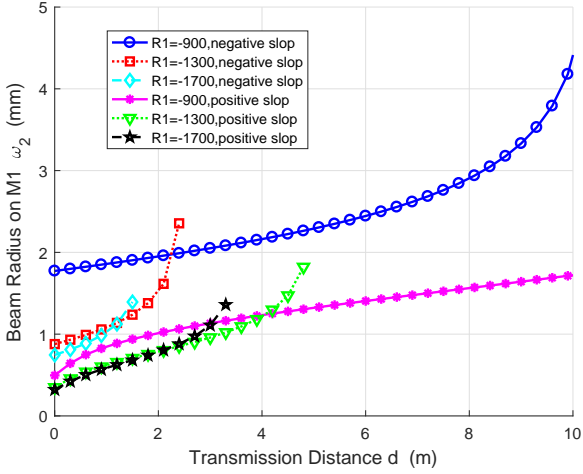}
	\caption{Beam Radius on M1 vs. Transmission Distance}

	\label{fig:beammirror1}
\end{figure}

\begin{figure}
	\centering
	\includegraphics[scale=0.6]{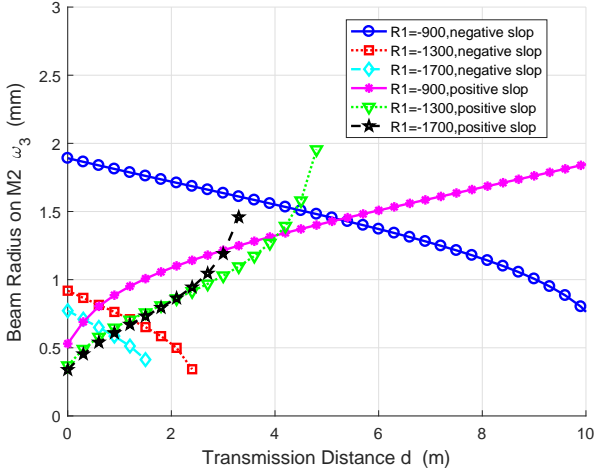}
	\caption{Beam Radius on M2 vs. Transmission Distance}
	\label{fig:beammirror}
\end{figure}
Therefore, the beam transmission distance $d$ can take any value lower than $d_{max}$. However, with the variation of $d$, the beam radiuses on the gain medium $\omega_1$, on M1 $\omega_2$, and on M2 $\omega_3$ take different value. Based on \eqref{equ:g1}, \eqref{equ:g2}, \eqref{omega12}, \eqref{omega22}, and \eqref{omega32}, how $\omega_1$, $\omega_2$, $\omega_3$ change as $d$ varies are given in Fig.~\ref{fig:beammedium}, Fig.~\ref{fig:beammirror1} and Fig.~\ref{fig:beammirror}, respectively.

As can be seen, in the case of positive slope, the beam radiuses on gain medium $\omega _1$, on M1 $\omega _2$ and on M2 $\omega _3$ increase with the increment of transmission distance. However, in the case of negative slope, the beam radiuses $\omega _1$, $\omega _2$ increase with the increment of transmission distance, while the beam radius $\omega _3$ is the opposite. At the same time, if with same $R1$, the beam radiuses $\omega _1$ and $\omega _2$ are smaller in the case of positive slope than that of negative slope.

The size of the beam radius on gain medium determines the minimum size of the gain medium, and the cost of the gain medium is determined by the size. In addition, the size of the gain medium will also affect the transmitter's size. For practical applications, the transmitter should be as smaller as possible. Given certain $R1$, to guarantee longer transmission distance and reduce the size of the transmitter and the cost of the gain medium, we chose the stability region with the positive slope.

For practical applications, we expect that the maximum transmission distance of the RBC system is at least $5m$. According to Fig.~\ref{fig:transdistance}, the $R1$ should satisfy $- 1300mm \le R1 \le  - 900mm$ in the case of positive slope.
\subsubsection{Transmission Efficiency Modeling}
After the above analysis, the transmission distance and the structure of the resonant cavity are determined.
\begin{figure}
	\centering
	\hspace{-9mm} \includegraphics[width=9.3cm,height=6cm]{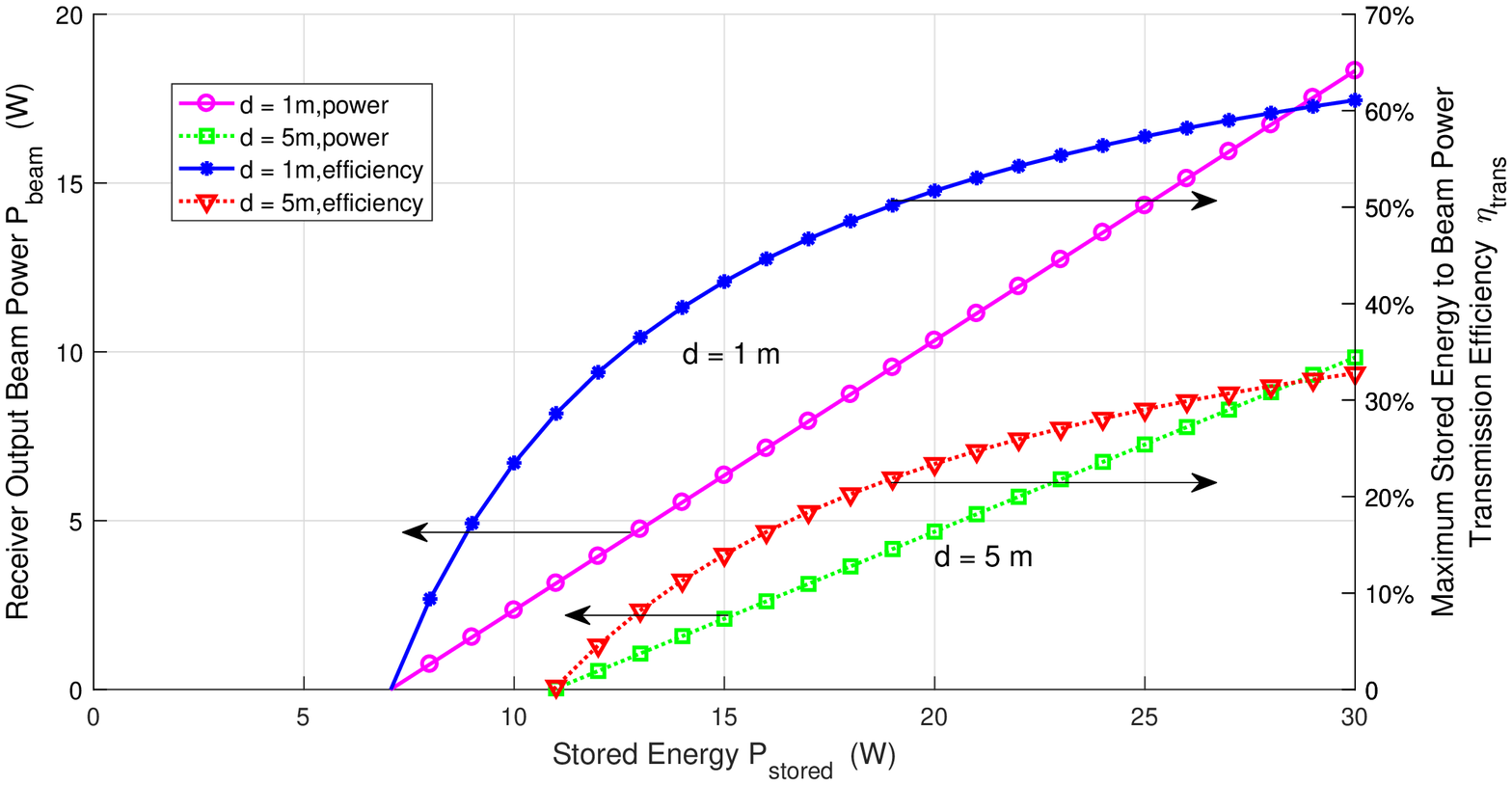}
	\caption{Beam Power and Transmission Efficiency vs. Stored Energy}
	\label{pbeampstore}
\end{figure}
As in \cite{SolidLaser}, $R$ of the mirror M2 in the receiver is $0.88$, the overlap efficiency $m$ is $1$ and $C$ can be $-5.64$. From \eqref{gs-pbeampstore} and \eqref{gs-fddelta}, given $d$, the function $f(d)$ becomes a constant, so $P_{beam}$ linearly depends on $P_{stored}$. To explore the relationship between $P_{beam}$ and $P_{stored}$, we set $d$ with $1m$ and $5m$. The relationships are shown as the linear lines in Fig.~\ref{pbeampstore}. As can be seen, when $P_{stored}$ is over the threshold, $P_{beam}$ increases as $P_{stored}$ increases. On the other hand, the threshold goes up if $d$ is larger. This is because that more stored energy attenuates when passing over a longer distance. Therefore, for certain $P_{stored}$, $P_{beam}$ takes smaller value when $d$ takes bigger value.

For the same $d$, the stored energy to the beam power conversion efficiency $\eta_{trans}$ can be obtained according to \eqref{gs-etatrans}. The relationship between $\eta_{trans}$ and $P_{stored}$ are given as the non-liner curves in Fig.~\ref{pbeampstore}. After $P_{stored}$ is over the threshold, $\eta_{trans}$ goes up dramatically at first, and then its growing trend becomes slow. At the same time, $\eta_{trans}$ takes smaller value if the beam goes through a longer distance. If the transmitting distance is $1m$, $\eta_{trans}$ can be up to about $61\%$ when $P_{stored}$ is $30W$.

\begin{figure}
	\centering
    \hspace{-9mm}
	\includegraphics[width=9.3cm,height=6cm]{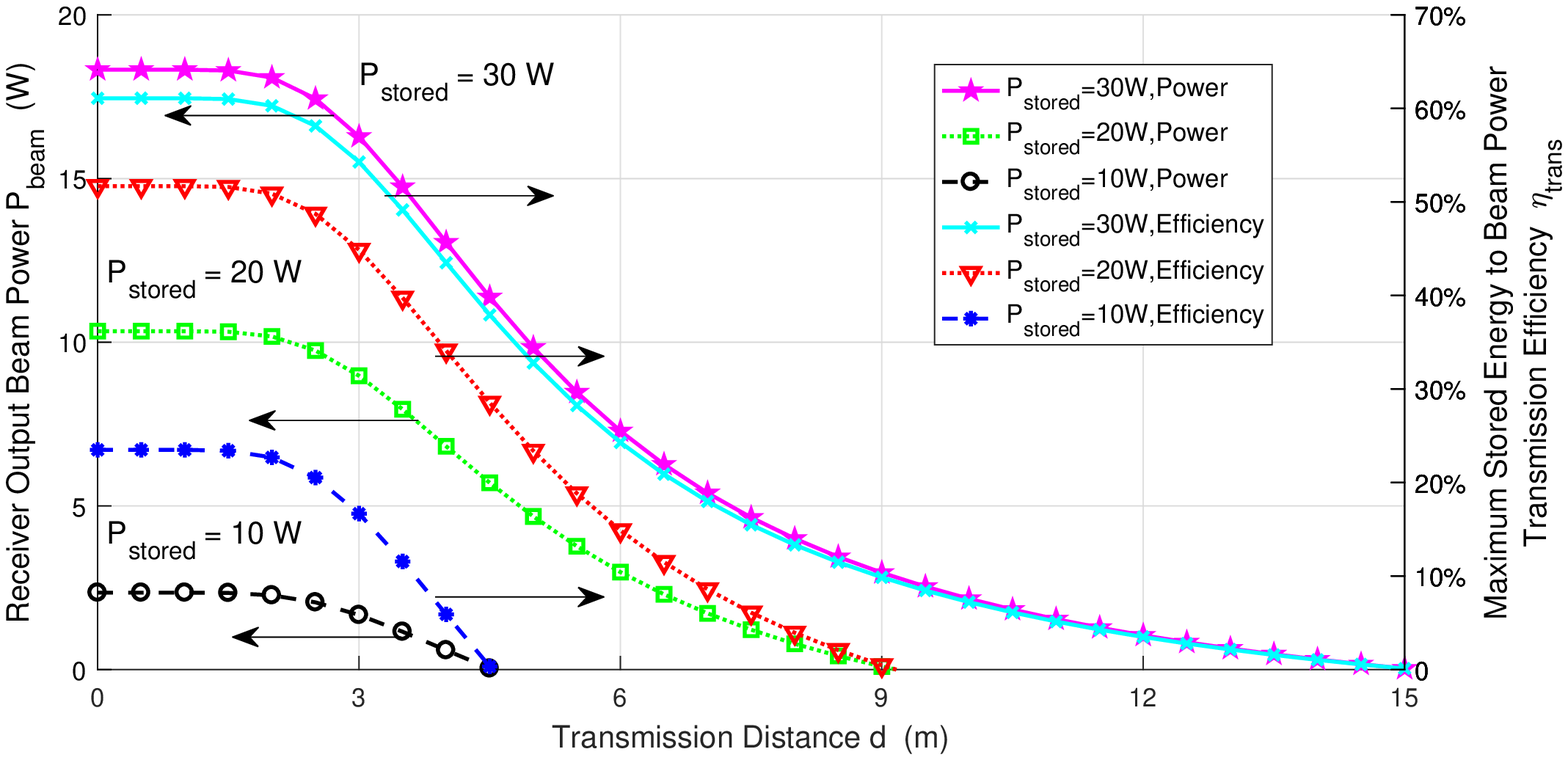}
	\caption{Beam Power and Transmission Efficiency vs. Transmission Distance}
	\label{pbeamd}

\end{figure}

From \eqref{gs-pbeampstore}, if $P_{stored}$ is given, with reference to \eqref{gs-etatrans}, the function $f(d)$ varies with the transmitting distance $d$, and $P_{beam}$ relies on $d$. Curves in Fig.~\ref{pbeamd} depicts how $P_{beam}$ and $\eta_{trans}$ changes with $d$, when $P_{beam}$ takes $10W$, $20W$, and $30W$, respectively. When the $d$ takes small values at first, $P_{beam}$ keeps almost steady. Then, as $d$ goes up, $P_{beam}$ goes down. Compared with the downward trend of $P_{beam}$ with low value, the downward trend of $P_{beam}$ with high value is much slower. Meanwhile, if with same $d$, higher $P_{stored}$ guarantees higher $P_{beam}$ being received at the receiver.

While, the changing trend of $\eta_{trans}$ over $d$ is similar with that of $P_{beam}$ when $P_{stored}$ takes same value. At the same time, for same $d$, $\eta_{trans}$ takes higher values when $P_{stored}$ takes higher value. For example, if $d$ is $3m$, $\eta_{trans}$ is more than $50\%$ when $P_{stored}=30W$, while $\eta_{trans}$ is about $40\%$ when $P_{stored}=20W$, and $\eta_{trans}$ only takes lower than $20\%$ when $P_{stored}=10W$. In the gain medium stored energy, part of the energy that is not converted to resonant beam is scattered into the surrounding environment. Safety can be guaranteed because the scattered power is with a very low power density.

 \subsection{Beam Power-to-Electrical Power Conversion}
At the RBC receiver, the PV-panel takes the role of converting the external-cavity beam power to the electrical power. Here we use the $808nm$ laser diode side pumped $1064nm$ Nd:YAG rod to establish the RBC system with $2m$ transmission distance, and record all the measured data. After analyzing the data with Simulink, we can obtain the value of $a_1$ and $b_1$ in \eqref{gs-pmpppbeam}, which are $0.3487$ and $-1.535$ respectively.

\begin{figure}
	\centering
\hspace{-6mm}
	\includegraphics[width=8cm,height=6cm]{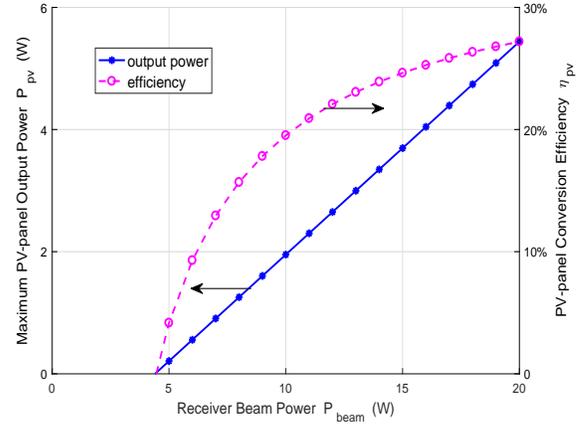}
	\caption{PV-panel Output Power and Conversion Efficiency vs. Beam Power}
	\label{ppvpbeam}
\end{figure}

Therefore, we can obtain the relationship between the maximum PV-panel output power $P_{pv}$ and the received external-cavity beam power $P_{beam}$, which is shown as the linear line in Fig.~\ref{ppvpbeam}. As only after accumulating a certain amount of energy can the PV-panel converts the external-cavity beam power to the electrical power, there is a threshold before the electrical power is output. After $P_{beam}$ becomes larger than the threshold, $P_{pv}$ increases over $P_{beam}$.

Then, the relationship between the PV-panel efficiency $\eta_{pv}$ and $P_{beam}$ can be obtained. The non-linear curve in Fig.~\ref{ppvpbeam} presents the relationship. After $P_{beam}$ is larger than the threshold, $\eta_{pv}$ goes up dramatically as $P_{beam}$ increases. Then $\eta_{pv}$ increases more slowly, and the maximum value of $\eta_{pv}$ is about $27\%$.
\subsection{End-to-end Power Transmission}

\begin{figure}
	\centering
	\includegraphics[width=8cm,height=6cm]{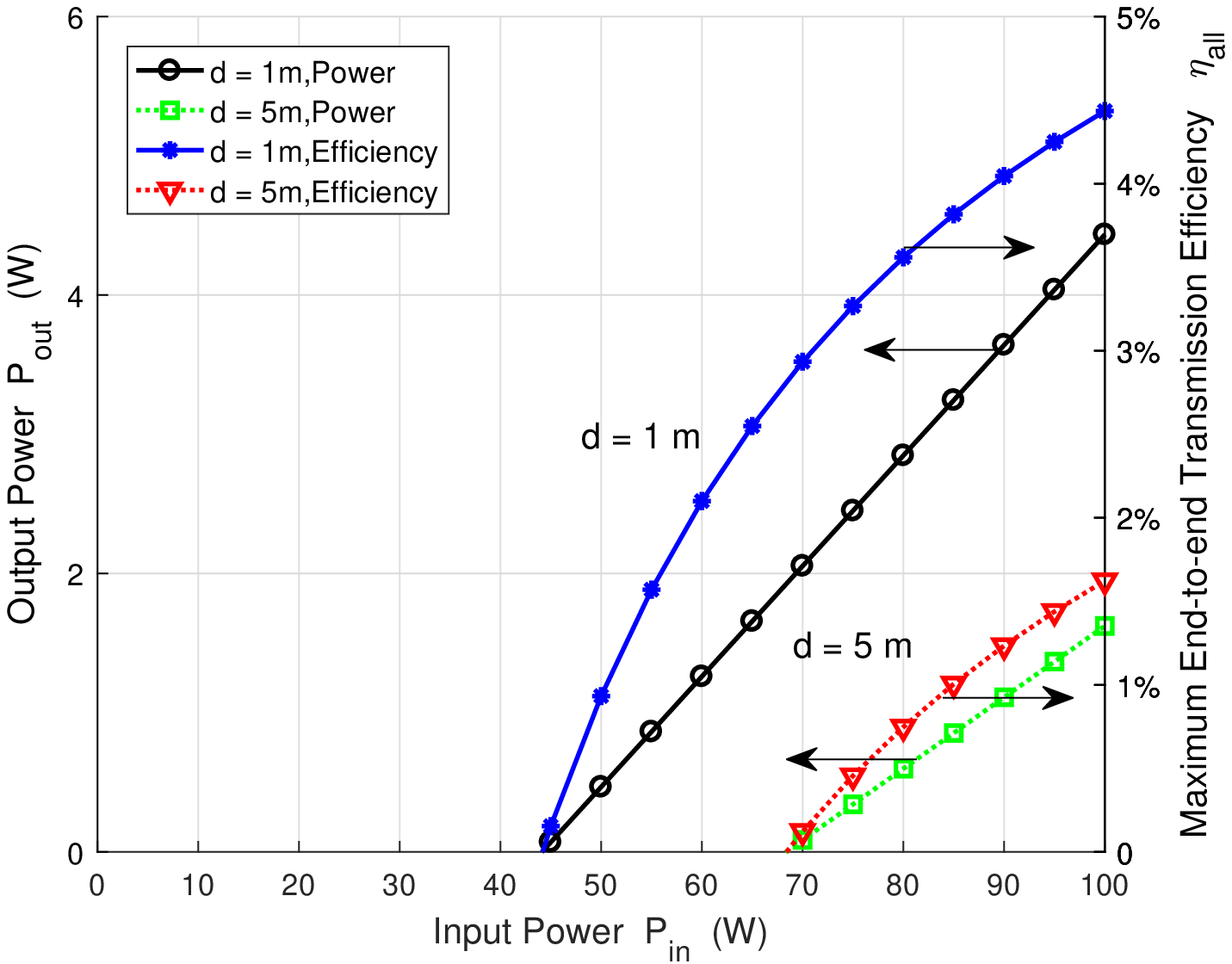}
	\caption{Output Power and Transmission Efficiency vs. Input Power}
	\label{poutpin}
	\centering
	\includegraphics[width=8cm,height=6cm]{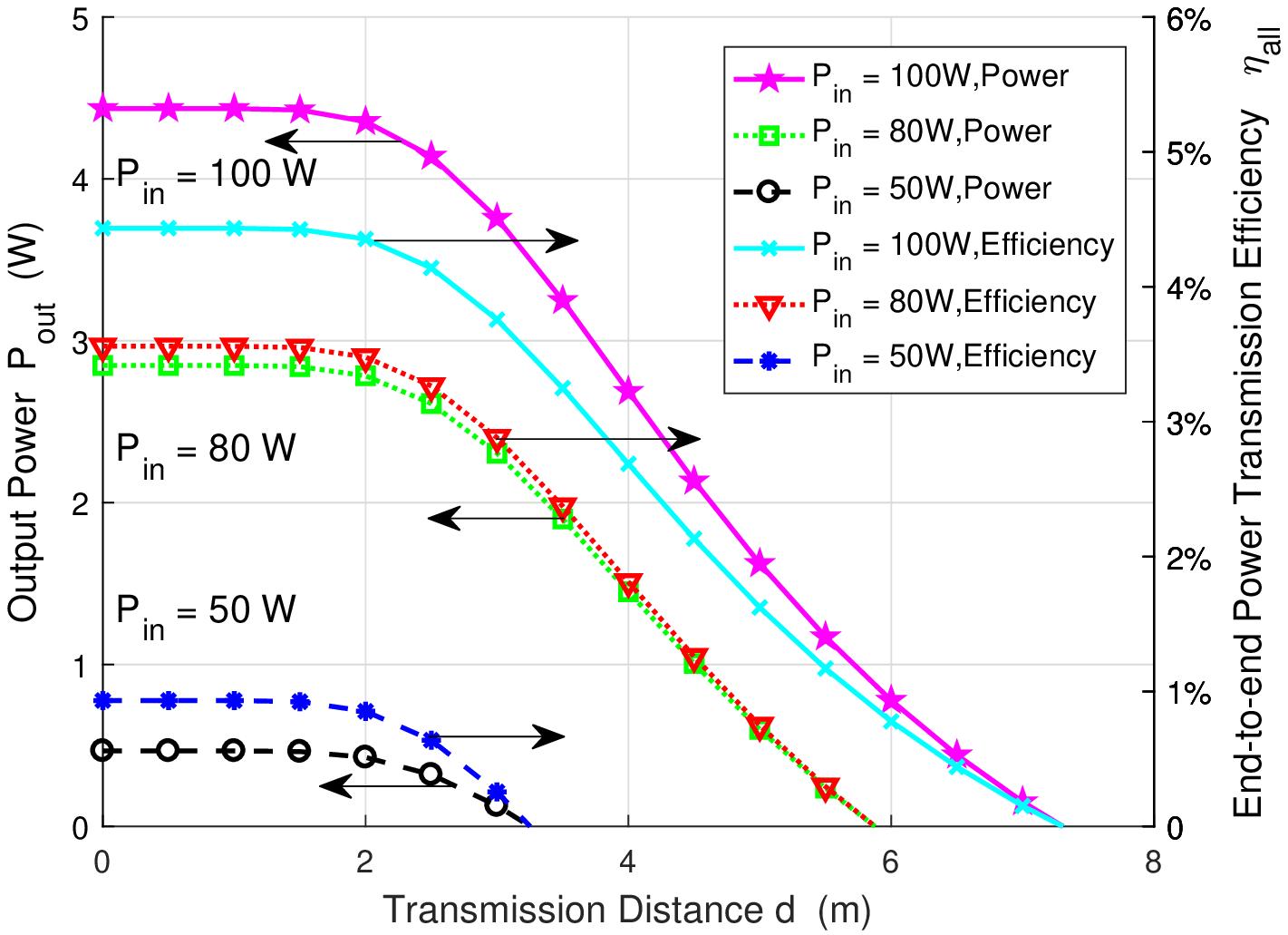}
	\caption{Output Power and Transmission Efficiency vs. Transmission Distance}
	\label{poutd}
\end{figure}

Based on the above analysis, the end-to-end power transmission procedure, that is the whole RBC charging system energy transmission procedure, can be depicted. The end-to-end power relationship, that is the relationship between the input electrical power for driving the system and the output electrical power for charging devices, can be obtained. Based on \eqref{pe2e}, when the transmission distance $d$ is given, the end-to-end power takes a linear relationship, which can be depicted as the linear lines in Fig.~\ref{poutpin}. As stated above, there are thresholds of the stored energy to the beam power transmission and the beam power to the electrical power conversion, and there is a threshold in the end-to-end power conversion relationship. After exceeding the threshold, $P_{out}$ goes up linearly as $P_{in}$ increases. At the same time, if with same $P_{in}$, the shorter distance guarantees the bigger $P_{out}$.

Then, from \eqref{etae2e}, how the end-to-end power transmission efficiency $\eta_{all}$ changes over $P_{in}$ can be obtained. The non-linear curves in Fig.~\ref{poutpin} depict the relationship between $\eta_{all}$ and $P_{in}$. As can be seen, $\eta_{all}$ goes up gradually with the increment of $P_{in}$ after that $P_{in}$ is over the threshold. From Fig.~\ref{poutpin}, we can obtain how much input power should be provided if we want to charge devices with certain power at certain distance. For example, if we want to charge our devices with $1W$, when the distance form our devices to the RBC transmitter is $1m$, the input power to the transmitter should be about $55W$; while if the distance is $5m$, the input power should be about $85W$. At the same time, the corresponding end-to-end efficiency can be obtained under each condition as well.

On the other hand, for certain $P_{in}$, the changing trend among $P_{out}$, $\eta_{all}$ and the transmission distance $d$ can be depicted. Fig.~\ref{poutd} gives all the relationships when  $P_{in}$ takes $50W$, $80W$, and $100W$. $P_{out}$ keeps almost steady when $d$ is relatively short. Then, with the increment of $d$, $P_{out}$ goes down dramatically. For different $P_{in}$, the declination takes different magnitude. The magnitude of the declines in $P_{out}$ when  $P_{in}$ takes $100W$ is slower than that when $P_{in}$ takes $80W$, which is slower than that when $P_{in}$ takes $50W$.

The changing law of $\eta_{all}$ over $d$ has similar trend with that of $P_{out}$ when $P_{in}$ takes same value. $\eta_{all}$ keeps almost steady at first and then decreases dramatically as well. Meanwhile, the RBC charging system is more efficient when $P_{in}$ takes larger value. When $P_{in}=100W$, the maximum $\eta_{all}$ is about $4.5\%$.

Fig.~\ref{poutd} provides guidelines for power control, efficiency control and distance selection when designing the RBC system. For example, if we want to support charging devices within $2m$, and design a system with about $4\%$ efficiency, we should provide $100W$ input power or about $3.6W$ output power. If we want to charge devices with $1.5W$ power, and the transmission distance should be as long as $4m$, then the preferred input power should be $80W$, and the end-to-end efficiency is about $3\%$.
In summary, the factors influencing the overall RBC efficiency include the input electrical power, the electrical power to stored energy conversion efficiency, the transmission distance and the PV-panel conversion efficiency. While the maximum transmission distance is decided by the stability conditions of the optical resonant cavity, the material and size of the gain medium, the diameter of the transmitter and the receiver, and the diffraction loss. We can obtain the following information from the above analysis:
\begin{itemize}
  \item We obtain the maximum energy transmission distance $d$ for the RBC system.
  \item We analyze the factors that influence the energy transmission power and efficiency, and gives their attenuation model over $d$.
  \item We derive the end-to-end RBC energy transmission efficiency $\eta_{all}$ in closed-form.
  \item We analyze the relationships among $P_{out}$, $\eta_{all}$, $P_{in}$, and $d$ under different circumstances, which lead to the design and development guidelines for the RBC system.
\end{itemize}


\section{Conclusions}
\label{sec:conclusion}
To provide perpetual power supply to IoT devices, resonant beam charging (RBC) is a promising safe, long-range and high-power wireless power transfer solution. In this paper, we present the analytical model of the energy transmission channel for the resonant beam charging (RBC) system, and study how long distance RBC can reach and how much power RBC can transfer. Based on the energy transmission channel modeling, we illustrate the RBC system performance influenced by the transmission distance, the transmission efficiency, and the output electrical power. These analysis lead to the guidelines for the RBC system design and implementation, which can deliver multi-Watt power over multi-meter distance wirelessly for IoT devices.

There are some open issues to be studied in future work. For example:
\begin{itemize}
  \item Since there are other loss factors influencing the RBC efficiency in addition to the diffraction loss. How to improve the overall efficiency of the RBC system should be further investigated.
  \item The RBC system performance will also be affected by other factors such as temperature. Thus the stability and reliability of the RBC system need further study.
\end{itemize}

\bibliographystyle{IEEEtran}
\bibliographystyle{unsrt}
\bibliography{references}

\begin{thebibliography}{10}
\providecommand{\url}[1]{#1}
\csname url@samestyle\endcsname
\providecommand{\newblock}{\relax}
\providecommand{\bibinfo}[2]{#2}
\providecommand{\BIBentrySTDinterwordspacing}{\spaceskip=0pt\relax}
\providecommand{\BIBentryALTinterwordstretchfactor}{4}
\providecommand{\BIBentryALTinterwordspacing}{\spaceskip=\fontdimen2\font plus
\BIBentryALTinterwordstretchfactor\fontdimen3\font minus
  \fontdimen4\font\relax}
\providecommand{\BIBforeignlanguage}[2]{{%
\expandafter\ifx\csname l@#1\endcsname\relax
\typeout{** WARNING: IEEEtran.bst: No hyphenation pattern has been}%
\typeout{** loaded for the language `#1'. Using the pattern for}%
\typeout{** the default language instead.}%
\else
\language=\csname l@#1\endcsname
\fi
#2}}
\providecommand{\BIBdecl}{\relax}
\BIBdecl

\bibitem{huang2013iot}
Y.~Zhou, C.~Huang, T.~Jiang, and S.~Cui, ``Wireless sensor networks and the
  internet of things: Optimal estimation with nonuniform quantization and
  bandwidth allocation,'' \emph{IEEE Sensors Journal}, vol.~13, no.~10, pp.
  3568--3574, Oct. 2013.

\bibitem{ling2017iot}
T.~Chen, Y.~Shen, Q.~Ling, and G.~Giannakis, ``Online learning for
  thing-adaptive fog computing in {IoT},'' in \emph{ASILOMAR}, Pacific Grove,
  USA, Oct. 2017.

\bibitem{ding2014iot}
Q.~Wu, G.~Ding, and Y.~Xu, ``Cognitive internet of things: {A} new paradigm
  beyond connection,'' \emph{IEEE Internet of Things Journal}, vol.~1, no.~2,
  pp. 129--143, Apr. 2014.

\bibitem{yu2017bigdata}
S.~Yu, M.~Liu, W.~Dou, X.~Liu, and S.~Zhou, ``Networking for {B}ig {D}ata: {A}
  survey,'' \emph{IEEE Communications Surveys and Tutorials}, vol.~19, no.~1,
  pp. 531--549, 2017.

\bibitem{Carroll2010}
A.~Carroll and G.~Heiser, ``An analysis of power consumption in a smartphone,''
  in \emph{Proceedings of the 2010 USENIX Conference on USENIX Annual Technical
  Conference}.\hskip 1em plus 0.5em minus 0.4em\relax Berkeley, CA, USA: USENIX
  Association, Jun. 2010, pp. 21--21.

\bibitem{Georgiou2017}
K.~Georgiou, S.~{Xavier-de-Souza}, and K.~Eder, ``The {IoT} energy challenge: A
  software perspective,'' \emph{IEEE Embedded Systems Letters}, pp. 1--4, Aug.
  2017.

\bibitem{Yu2016A}
S.~Yu, W.~Zhou, S.~Guo, and M.~Guo, ``A feasible {IP} traceback framework
  through dynamic deterministic packet marking,'' \emph{IEEE Transactions on
  Computers}, vol.~65, no.~5, pp. 1418--1427, May. 2016.

\bibitem{zhang2017networks}
H.~Zhang, Y.~Qiu, X.~Chu, K.~Long, and V.~C.M.Leung, ``Fog radio access
  networks: {M}obility management, interference mitigation, and resource
  optimization,'' \emph{IEEE Wireless Communications}, vol.~24, no.~6, pp.
  120--127, Dec. 2017.

\bibitem{wirelesstechniques}
X.~Lu, D.~Niyato, P.~Wang, D.~I. Kim, and Z.~Han, ``Wireless charger networking
  for mobile devices: {F}undamentals, standards, and applications,'' \emph{IEEE
  Wireless Communications}, vol.~22, no.~2, pp. 126--135, Apr. 2015.

\bibitem{electromagnetic}
A.~Costanzo, M.~Dionigi, D.~Masotti, M.~Mongiardo, G.~Monti, L.~Tarricone, and
  R.~Sorrentino, ``Electromagnetic energy harvesting and wireless power
  transmission: {A} unified approach,'' \emph{Proceedings of the IEEE}, vol.
  102, no.~11, pp. 1692--1711, 2014.

\bibitem{inductive}
S.~Valtchev, E.~Baikova, and L.~Jorge, ``Electromagnetic field as the wireless
  transporter of energy,'' \emph{Facta universitatis - series: Electronics and
  Energetics}, vol.~25, no.~3, pp. 171--181, Dec. 2012.

\bibitem{magnetic}
S.~Assawaworrarit, X.~Yu, and S.~Fan, ``Robust wireless power transfer using a
  nonlinear parity-time-symmetric circuit,'' \emph{Nature}, vol. 546, no. 7658,
  p. 387, Jun. 2017.

\bibitem{Cannon2009}
B.~Cannon and J.~Hoburg, ``Magnetic resonant coupling as a potential means for
  wireless power transfer to multiple small receivers,'' \emph{IEEE
  Transactions on Power Electronics}, vol.~24, no.~7, pp. 1819--1825, Jul.
  2009.

\bibitem{RF}
Z.~Popovic, ``Cut the cord: Low-power far-field wireless powering,'' \emph{IEEE
  Microwave Magazine}, vol.~14, no.~2, pp. 55--62, Apr. 2013.

\bibitem{Lasercharging}
J.~Fakidis, S.~Videv, H.~Helmers, and H.~Haas, ``0.5-{Gb/s} {OFDM}-based laser
  data and power transfer using a {GaAs} photovoltaic cell,'' \emph{IEEE
  Photonics Technology Letters}, pp. 1--1, Mar. 2018.

\bibitem{Sahai2011}
A.~Sahai and D.~Graham, ``Optical wireless power transmission at long
  wavelengths,'' in \emph{2011 International Conference on Space Optical
  Systems and Applications (ICSOS)}, Santa Monica, CA, USA, May 2011, pp.
  164--170.

\bibitem{liu2016dlc}
Q.~Liu, J.~Wu, P.~Xia, S.~Zhao, W.~Chen, Y.~Yang, and L.~Hanzo, ``Charging
  unplugged: {W}ill distributed laser charging for mobile wireless power
  transfer work?'' \emph{IEEE Vehcular Technology Magzine}, vol.~11, no.~4, pp.
  36--45, Dec. 2016.

\bibitem{Qing2017}
Q.~Zhang, X.~Shi, Q.~Liu, J.~Wu, P.~Xia, and Y.~Liao, ``Adaptive distributed
  laser charging for efficient wireless power transfer,'' in \emph{2017 IEEE
  86th Vehicular Technology Conference (VTC2017-Fall)}, Toronto, ON, Canada,
  Sep. 2017, pp. 1--5.

\bibitem{SolidLaser}
W.~Koechner, \emph{{S}olid-{S}tate {L}aser {E}ngineering}, 6th~ed.\hskip 1em
  plus 0.5em minus 0.4em\relax Springer, 2006.

\bibitem{diffraction}
S.~Cao, S.~Tu, Y.~Huang, H.~Fan, J.~Li, H.~Xia, and G.~Ren, ``The analysis in
  the resonator diffraction losses of the laser modes,'' \emph{Laser
  Technology}, vol.~42, no.~3, pp. 400--403, Sep. 2018.

\bibitem{secondauthorPV}
M.~S. Aziz, S.~Ahmad, and I.~Husnaini, ``Simulation and experimental
  investigation of the characteristics of a {PV}-harvester under different
  conditions,'' in \emph{International Conference on Energy Systems and
  Policies}, Nov. 2014, pp. 1--8.

\bibitem{green2015solar}
A.~G. Martin, E.~Keith, H.~Yoshihiro, W.~Wilhelm, and D.~D. Ewan, ``Solar cell
  efficiency tables ({V}ersion 45),'' \emph{Progress in photovoltaics: research
  and applications}, vol.~23, no.~1, pp. 1--9, 2015.

\bibitem{Summerer2008Concepts}
L.~Summerer and O.~Purcell, ``Concepts for wireless energy transmission via
  laser,'' in \emph{International Conference on Space Optical Systems and
  Applications}, 2008.

\bibitem{LENS}
H.~Kogelnik, ``{I}maging of optical modes - resonators with internal lenses,''
  \emph{The Bell System Technical Journal}, vol.~44, no.~3, pp. 455--494, Mar.
  1965.

\bibitem{stable}
H.~Kogelnik and T.~Li, ``{L}aser {B}eams and {R}esonators,'' \emph{Applied
  Optics}, vol.~5, no.~10, pp. 1550--1567, Oct. 1966.

\bibitem{stable0}
V.~Magni, ``{R}esonators for solid-state lasers with large-volume fundamental
  mode and high alignment stability,'' \emph{Applied Optics}, vol.~25, no.~1,
  pp. 107--117, Jan. 1986.

\bibitem{Laser}
B.~Zhou, Y.~Gao, and J.~Chen, \emph{{L}aser {P}rinciple}, 6th~ed.\hskip 1em
  plus 0.5em minus 0.4em\relax Beijing: The National Defense Press, 2004.

\bibitem{Loss}
G.~Wei and B.~Zhu, \emph{{L}aser {B}eam {O}ptics}, 1st~ed.\hskip 1em plus 0.5em
  minus 0.4em\relax Beijing: Beijing Industry Collage Press, 1988.

\end{thebibliography}

\end{document}